\documentstyle[preprint,prbbib,aps]{revtex}
\begin{document}
\draft
\title{Current density functional  theory of quantum dots in a
magnetic field.}
\author{M. Ferconi and G. Vignale}
\address{Institute for Theoretical Physics,
University of California, Santa Barbara, California 93106-4030 \\
and \\
Department of Physics and Astronomy,
University of Missouri-Columbia,
Columbia, Missouri 65211}
\date{\today}
\maketitle
\begin{abstract}
We present a study of ground state energies and densities of quantum dots
in a magnetic field, which takes into account correlation effects
through the Current-density functional theory (CDFT). The method is
first tested against exact results  for the energy and density of 2 and
3 electrons quantum dots, and it is found to yield an accuracy better
than $ 3 \%. $ Then we extend the study  to larger dots and compare
the results with available experimental data.
The orbital and spin angular momenta of the ground state,
and the evolution of the density profile as a function of the
magnetic field are calculated. Quantitative evidence of edge reconstruction
at high magnetic field is presented.
\end{abstract}
\narrowtext

Recent progress  in nanofabrication techniques has   allowed the realization
of microstructures in which electrons  in a two dimensional electron gas
are confined to an approximately circular region, of radius varying from
several hundreds to several thousand Angstroms. At the densities
characteristic of these quantum dots $(n \sim  10^{11} \div 10^{12}
{\rm cm}^{-2})$, as a consequence of the material that hosts the
system, a magnetic field $ B $ of the order of few Tesla is already
``strong'', i.e., it has
a magnetic length $ \lambda = \sqrt{\hbar c / e B} $  of the order of the
interelectron distance. For this reason, particular attention has been
devoted to the evolution of the ground state properties as function of
an applied magnetic field.
Experiments using conductance~\cite{mceuen1},
and single-electron capacitance~\cite{ashoori} spectroscopy have
led to the measurement of the electronic spectrum of the dot, and
of its addition energy, namely the energy necessary to add one electron
to the system, as a function of the magnetic field.  The occurence of
frequent cusps in these functions has been interpreted in terms of
crossings between energy levels characterized by different values of
spin and/or orbital angular momentum.  In the strong magnetic
field regime, interesting effects, such as  ``magic'' angular momentum
quantum numbers, formation of Wigner molecules, and edge reconstruction,
have been theoretically predicted.~\cite{maks_phys,chamon1}

{}From the standpoint of the theory, the main obstacle to the determination
of the energy level structure of quantum dots, is the strong
electron-electron interaction -- the non-interacting problem being
exactly solvable, at least in the approximation of parabolic
confinement.~\cite{fock-darwin}
Exact diagonalization studies have been carried out for dots with a
small number of electrons~\cite{Maks-Chak,Pfann,Haw1,Yang} ($N \leq 8$),
and have yielded a very rich scenario.
In the case of larger systems, one has to resort to some approximations
in order to carry on the calculations. Approximation schemes such as the
constant interaction model~\cite{beenhk},
self-consistent  Thomas-Fermi method,~\cite{mceuen2} the
Hartree-Fock,~\cite{chamon1,palacios} or other mean-field
approximations,~\cite{fogler} all have  the common feature of  not
including correlation effects.  The importance  of the latter is all but
negligi
when the number of electrons in the dot is
less than  few hundreds.
For the unpolarized two electrons case, correlation effects make up to
$\sim 10\% $ of the energy,~\cite{Pfann}  which is quite a sizeable
quantity. At the same time the corresponding Hartree-Fock density
profile is incorrect. Moreover, the magnetic field values  at which
changes in the  quantum numbers occur are  influenced by correlation
effects.  All the above are  compelling reasons for  including
correlations in any model that wants to describe quantitatively
the properties of quantum  dots.

 A full fledged many-body approach
appears  to be  prohibitively difficult at present, for dots containing
more than few electrons.
On the other hand, density functional theory~\cite{GrossDreizler} (DFT)
is known to  have been successfully applied  to many atomic and molecular
systems, and it provides -- via the local density approximation (LDA) -- a
simple method to approximately include electronic correlations.
In the present paper, we employ  the generalization of DFT known as
Current Density Functional Theory,~\cite{cdft} which is appropriate
to deal with systems in the presence of a magnetic field. We emphasize
that this theory had never been previously
tested against {\it exact} results. Quantum dots with only few electrons have
enabled us to conduct such a test for the first time.

The CDFT scheme for electronic systems in a magnetic field has been
reviewed extensively.~\cite{cdft} Here we limit ourselves to the essentials
of the formulation for quantum dots. We consider circularly symmetric dots,
and assume a parabolic confining potential
$ V(r) = 1/2 \; m \omega_{0}^{2} r^{2}$
of frequency $\omega_{0},$ with $ m $ the effective electron mass.
The vector potential corresponding to the uniform magnetic field in the
symmetric gauge is
$ {\bf A}({\bf r}) =  B / 2 (-y,x,0) \label{Asymgau}  . $
The ground state energy $E,$ the density $ n({\bf r}),$ and the paramagnetic
current density ${\bf j}_{p} ({\bf r}),$ are all expressed in terms of a
set of Kohn-Sham orbitals
\begin{equation}
\psi_{j l \sigma }({\bf r}) = e^{-i l \theta}  \phi_{j l \sigma}(r) ,
\end{equation}
which are eigenstates of the $ z $ component of the angular momentum $-l$,
and satisfy the Kohn-Sham equation
\begin{eqnarray}
\Biggl\{
- \frac{\hbar^{2}}{2 m } \left( \frac{\partial^2}{\partial r^{2}} +
\frac{1}{r} \frac{\partial}{\partial r} - \frac{l^{2}}{r^{2}} \right) -
\frac{e \hbar l}{2 m c} B +  \frac{m \Omega^{2}}{2} r^{2} & &
\Biggr. \nonumber \\
 -  \Biggl. \frac{e \hbar l }{m c } \frac{A_{xc}}{r} + V_{H} +
V_{xc \; \sigma}
+ g \mu_{B} B  \sigma \Biggr\}     \phi_{i l \sigma} & = &
\varepsilon_{i l \sigma} \phi_{i l \sigma} , \label{finKS}
\end{eqnarray}
$\sigma$ being  the  $z$ component of the spin and
$\mu_{B} = e \hbar / 2 m c $ the effective Bohr magneton.
The renormalized frequency  $ \Omega = \sqrt{\omega_{0}^{2} + \omega_{c}^{2}
/{4
where $\omega_{c}$ is  the cyclotron frequency.
Explicit expressions for the relevant quantities are
\begin{equation}
n_{\sigma}(r) = \sum_{\{il \}}^{N_{\sigma}} \;
| \phi_{i l \sigma}(r) |^{2} ,
\label{finden}
\end{equation}
for the spin density; ($\sigma = \{ \uparrow, \downarrow \}$),
\begin{equation}
{\bf j}_{p}(r) =  j_{p}(r) \hat{{\bf e}}_{\theta} =
- \frac{\hbar }{m r} \sum_{\{ i l \sigma \}}^{N} \; l \;
|  \phi_{i l \sigma}(r) |^{2}  \;
 \hat{{\bf e}}_{\theta} , \label{fincurp}
\end{equation}
for the paramagnetic current density,
with $ \hat{{\bf e}}_{\theta}$ the azimuthal unit vector,
\begin{eqnarray}
E & = & \sum_{\{ i l \sigma \} }^{N} \; \varepsilon_{i l \sigma}   -
\frac{e^{2}}{2 k}
\int \! \int \! d{\bf r}  \; d{\bf r}' \;
\frac{n({\bf r})  n({\bf r}')}{ |  {\bf r - r}' |} \nonumber \\
& - & \sum_{\sigma} \; \int \! d{\bf r} \; n_{\sigma}({\bf r}) \;
V_{xc \; \sigma}({\bf r})  - \frac{e}{c} \int \! d{\bf r} \;
{\bf j}_{p}({\bf r})  \cdot {\bf A}_{xc}({\bf r})  \nonumber \\
& + & E_{xc}[n,\xi,\mbox{\boldmath $\cal V $}] ,
\label{eqgren}
\end{eqnarray}
for the energy.
In all cases  the summation is extended to the  $N $ lowest
eigenvalues, where $ N $ is the number of electrons
($N = N_{\uparrow} + N_{\downarrow}$). The Kohn-Sham
equation includes the self-consistent Hartree potential
\begin{equation}
V_{H}(r) = 2 \pi \frac{e^{2}}{k} \int \! dr' \;
r' \frac{n(r')}{ |  {\bf r - r}' |} , \label{hartree}
\end{equation}
the exchange-correlation potential
\begin{equation}
V_{xc \: \sigma}({\bf r}) =
\left. \frac{\delta E_{xc}[n,\xi,\mbox{\boldmath $\cal V$}]}{\delta
n_{\sigma}({\bf r})}
\right|_{n_{- \sigma},\mbox{\boldmath $\cal V$}} -
\frac{e}{c} {\bf A}_{xc}({\bf r})
\cdot \frac{{\bf j}_{p}({\bf r})}{n({\bf r})} , \label{vxcCDFT}
\end{equation}
and the exchange-correlation vector potential
\begin{equation}
\frac{e}{c} {\bf A}_{xc} = \frac{e}{c} A_{xc}  \hat{{\bf e}}_{\theta} =
 \frac{m c }{e n(r)}  \frac{\partial}{\partial r}
\left. \left( \frac{\delta E_{xc}[n,\xi,
\mbox{\boldmath $\cal V $}]}{\delta \mbox{\boldmath $\cal V $}} \right)
\right|_{n,\xi} . \label{finaxc}
\end{equation}
These last two quantities are formally defined as functional derivatives
of the exchange correlation energy functional $E_{xc}$ -- a functional of the
density $ n, $ the spin polarization
$ \xi = (n_{\uparrow} - n_{\downarrow} ) / (n_{\uparrow} + n_{\downarrow} ), $
and the vorticity
$\mbox{\boldmath $\cal V $}({\bf r}) = {\cal V}(r) \hat{{\bf e}}_{z} =
- m c / e r  \; \partial \left[ r j_{p}(r) / n(r) \right] / \partial r
\;  \hat{{\bf e}}_{z}   .$
The exchange-correlation energy has been calculated in the local
density approximation (LDA)
\begin{equation}
E_{xc} = \int \! d{\bf r} \; n(r)
\epsilon_{xc}[n({\bf r}),\xi({\bf r}),\mbox{\boldmath $\cal V $}({\bf r})]
\label{ldaexc}
\end{equation}
where  $ \epsilon_{xc}$ is the exchange-correlation energy density
of the uniform two-dimensional electron gas in an effective magnetic field
$ \mbox{\boldmath $\cal V$}({\bf r})$ which, in the LDA is taken to be
approxima
equal to the external magnetic field ${\bf B}$.
Following  Rasolt and Perrot~\cite{RasPer} we have chosen
\begin{equation}
\epsilon_{xc}[n,\xi,\mbox{\boldmath $\cal V $}] =
\frac{\epsilon_{xc}^{LWM}[n,\nu] +
\nu^{4} \epsilon_{xc}^{TC}[n,\xi]}{1 + \nu^4} \label{endenexc}
\end{equation}
where $\nu= 2 \pi \lambda^{2} n $ is the filling factor.
The expression connects the fitted form of Levesque
{\em et al.,}~\cite{lwm} $ \epsilon_{xc}^{LWM}[n,\nu],$
which is valid for small filling factor (or large magnetic field),
to the form given by Tanatar and
Ceperley~\cite{TanCep} $ \epsilon_{xc}^{TC}[n,\xi] $  for zero magnetic field.
Different Pad\'{e}  polynomials   that reproduce the known
limiting cases do not change appreciably the general behavior in the
regions where they contribute to the energy.
It must be underlined that, in contrast
to previous approaches, the present method does not use a truncated
basis set of Landau or Fock-Darwin orbitals as a basis in which to
expand the solution of the Kohn-Sham equation. For this reason, our
approach is completely unbiased with respect to the strength of the
magnetic field and should work equally well, a priori, for weak and
strong magnetic fields.

The accuracy of  our method has been tested  in the two and three electrons
cases, where exact results are available.  This is a   very
severe test for  any  uniform electron gas based  energy functional.
Our  results are summarized in Fig.~(\ref{figgren2and3el}).
For two electrons, we plot the ground state energies
 of the $S = 0, $ and $ S= 1$
symmetry states, versus the  magnetic field.  It is seen that the deviation
from
is contained to $ \lesssim 3 \%  .$
A remarkable  improvement is achieved in the $ S = 0$
case upon the Hartree-Fock calculation~\cite{Pfann}, which gives an error of
$ \sim 10 \% $. Another satisfying feature, is that the orbital angular
momentum
to increase in  steps with increasing magnetic field - is
correctly obtained, apart from a slight difference in the value of the
magnetic field at which  the discontinuities occur.
Excellent agreement is also obtained in the ground-state density distributions
for each spin symmetry state
(see inset of Fig.\ref{figgren2and3el}a),
which again shows considerable improvement upon the
Hartree-Fock result for the singlet.
With increasing magnetic field,  the density distribution moves
inward and becomes more  localized about the origin,  as long
as the angular momentum remains zero.  This leads to an increase in
electrostatic energy which is suddenly released when the angular momentum
jumps to a  finite value, and the density peak moves outward to a finite
radius.  This pattern  is repeated  every time that the orbital angular
momentum increases.  The inward displacement of the density at constant
angular momentum, is followed by a sudden outward expansion  when the
angular momentum increases.

In the three electrons case,~\cite{Haw1}  we have plotted in
Fig.~(\ref{figgren2and3el}b)  the ground state
energies per particle  of the $S = 3/2 , $ and $ S = 1/2 $  ground states.
An energy of
$\hbar \Omega$ per particle has been subtracted in order to
show more clearly the rich structure of the  ground state energy, featuring
several changes in orbital angular momentum.  Even for this small part of the
total energy, CDFT is able to correctly reproduce the exact behavior, with an
error that even in the  worst situation is  below $ \sim 5 \%, $ being much
lower in general. As in the two electron case,  we note a loss of accuracy
for very high magnetic field,  due to the increase in the number of
quasi-degenerate Kohn-Sham orbitals. We also notice that
the exchange-correlation vector potential $A_{xc}$ gives a very small
contribution to the  energy.

The above results provide convincing evidence that CDFT is a powerful tool
to predict qualitatively - and quantitatively, within an accuracy of few
percent - the evolution of the ground state properties of quantum dots with
magnetic field.  We  next  apply it to the calculation of quantum dots with
larger number of electrons.  Figure~(\ref{figadden9el}), shows the ground state
 and orbital symmetries of 9 and 10 electrons quantum dots,
as well as the addition energy for the 9 electron system.
A comparison of the results with the experimental  data~\cite{ashoori}
is shown in the lowest panel of  Fig.~(\ref{figadden9el}).
Following Hawrylak~\cite{Haw2} we have assumed a dependence of the
effective confining potential  $ \omega_{N} $  on the number of  electrons $
N $ in the dot, of the form  $ \omega_{N} =  \sqrt{ \omega_{0}^{2} + e^{2} N
/ m k d^{3}}, $ where $ d $  is the distance of the positive charge
background from the plane of the dot, and  $\omega_{0}$ is the ``empty dot''
confining frequency.
A vertical rigid shift of the theoretical  curve has also
been performed on the basis that there is a constant term for the
energy at fixed number of  electrons in the dot,  depending on the geometry
of the environment in which the system is created and on the  probe that
measures the addition energies. Although the patterns of cusps in
the theoretical and experimental  curves appear to be very different,   we
suggest that the discrepancy may be attributed, in large part,  to  a
systematic  underestimation of the values of the magnetic field at which the
quantum numbers transitions occur.  In other words,  a rigid shift of
the theoretical curve to higher magnetic field would considerably
improve the agreement with experiment.    In general, however, the observed
difference suggests that the conventional model of a confinement  essentially
independent on  $N$ and $B$ may not be an adequate representation of
reality,  at the level of accuracy considered here.

As a final point of interest,  we have studied the evolution of the density
profile  as a function of magnetic field, and some results for the 10
electrons dot are presented in Fig.~(\ref{edgefig}).
According to   the classical electrostatic analysis~\cite{fogler},
the distribution should look like the dome shown in Fig.~(\ref{edgefig});
which is magnetic field independent,
with small corrections induced by the quantized kinetic energy.
While at low values of the magnetic field the electrostatic
behavior is reasonably reproduced, apart from oscillations, with increasing
magnetic field significant changes take place, reflecting the different
values of quantum numbers.  At $B  \simeq 3.9 T$ the  $10-$electrons
dot becomes fully spin polarized,  and its orbital angular momentum
$M_z=45$ corresponds to the ``maximum density droplet'' i.e, all
the orbitals with $l = 0,1...9$ are
occupied by a spin up electron.~\cite{MDD}  As the magnetic field is
further increased, some electrons are promoted to orbitals  with  $l > 9$
leaving some of the orbitals with $l < 9$  empty.  In terms of the density
profile, this effect appears as  an edge reconstruction,  similar to the one
discussed in Ref.~\onlinecite{chamon1},  whereby the charge density first
decreases and then increases  in the edge region, before falling to zero.
The physical reason is that the exchange-correlation energy
favors a higher density in the edge region,  while the electrostatic energy,
of course, opposes it.  The precise size and shape of the edge is determined
by the competition of these two energies. As it can be seen,
this  phenomenon is not restricted
to large dots, but occurs even in small systems.
More details on this problem will be published elsewhere.

We thank J.  Perdew for providing us an atomic density funtional code
from  which we have derived ours,  R. Ashoori for giving us
his experimental data, and J.\ Palacios for sending us a copy of his
work prior to publication.
We  acknowledge useful conversations with
Pawel  Hawrylak, Ray Ashoori, Eberhard Gross  and Francois Perrot.
We acknowledge support from NSF Grant No. DMR~9100988 and hospitality
at the ITP where part of this work was done, under NSF Grant No. PHY89-04035.

\begin{figure}
\caption{%
(a) Ground state energy for the  two electrons quantum dot,  for
$ S_{z} =0 $ and $S_{z}=1.$  The transition from
$(M_{z},2 S_{z}) = (0,0) $  to  $(M_{z},2 S_{z}) = (2,0) $ is marked
by an arrow.  In the inset the density profiles for the two spin
configurations  are plotted and compared with the exact
and Hartree-Fock (HF)~\protect\cite{Pfann} results.
(b) Ground state energy per electron
for  $S = 3/2,$ and
$ S = 1/2 $  in a three electrons quantum dot. The transition points from
orbital angular momentum $M_{z1}$ to $M_{z2}$  are indicated as
$M_{z1}  \rightarrow M_{z2}.$
Parameters appropriate for GaAs have
been chosen, i.e.  $ m = 0.067 m_{e},$ with $m_{e}$ electron mass, $k = 12.4,$
w
$a_{B}\sim 98 \AA,$  $ \omega_{0} = 3.37 meV. $
The Zeeman splitting is neglected.}
\label{figgren2and3el}
\end{figure}
\begin{figure}
\caption{(a)Magnetic field dependence of the ground state energy
for the 9 electrons dot. (b) Same for the 10 electrons dot. The quantum
numbers $(M_{z},2 S_{z}),$
of the ground states  for the 9 and 10 electrons are shown ,
and they refer to the  regions of magnetic field between two successive
arrows.
is a guide to the  eye.  A g factor of $g = -0.44,$ and
$\omega_{0} =  3.31 meV$ have been assumed.}
\label{figadden9el}
\end{figure}
\begin{figure}
\caption{Evolution of the density profile with the magnetic field
for the  10 electrons  quantum dot of Fig~(\protect\ref{figadden9el}). The
class
given by the dashed line, while the density corresponding to filling
factor $\nu = 1 = 1 / 2 \pi \lambda^{2}$ is represented by the
horizontal line. The quantum numbers $(M_{z},2 S_{z})$ are also shown.}
\label{edgefig}
\end{figure}

\end{document}